\documentclass{aip-cp}
\usepackage[numbers,elide]{natbib}
\usepackage{fancybox}
\usepackage{shortvrb}
\MakeShortVerb\|

\usepackage{listings}
\usepackage[usenames]{color}
\definecolor{codecolor}{gray}{.9}
\definecolor{rlcolor}{cmyk}{0,1,0,0}
\begin{document}

\let\jnl\rm
\def\refa@jnl#1{{\jnl#1}}

\def\aj{\refa@jnl{AJ}}                   
\def\actaa{\refa@jnl{Acta Astron.}}      
\def\araa{\refa@jnl{ARA\&A}}             
\def\apj{\refa@jnl{ApJ}}                 
\def\apjl{\refa@jnl{ApJ}}                
\def\apjs{\refa@jnl{ApJS}}               
\def\ao{\refa@jnl{Appl.~Opt.}}           
\def\apss{\refa@jnl{Ap\&SS}}             
\def\aap{\refa@jnl{A\&A}}                
\def\aapr{\refa@jnl{A\&A~Rev.}}          
\def\aaps{\refa@jnl{A\&AS}}              
\def\azh{\refa@jnl{AZh}}                 
\def\baas{\refa@jnl{BAAS}}               
\def\bac{\refa@jnl{Bull. astr. Inst. Czechosl.}}
\def\caa{\refa@jnl{Chinese Astron. Astrophys.}}
\def\cjaa{\refa@jnl{Chinese J. Astron. Astrophys.}}
\def\icarus{\refa@jnl{Icarus}}           
\def\jcap{\refa@jnl{J. Cosmology Astropart. Phys.}}
\def\jrasc{\refa@jnl{JRASC}}             
\def\memras{\refa@jnl{MmRAS}}            
\def\mnras{\refa@jnl{MNRAS}}             
\def\na{\refa@jnl{New A}}                
\def\nar{\refa@jnl{New A Rev.}}          
\def\pra{\refa@jnl{Phys.~Rev.~A}}        
\def\prb{\refa@jnl{Phys.~Rev.~B}}        
\def\prc{\refa@jnl{Phys.~Rev.~C}}        
\def\prd{\refa@jnl{Phys.~Rev.~D}}        
\def\pre{\refa@jnl{Phys.~Rev.~E}}        
\def\prl{\refa@jnl{Phys.~Rev.~Lett.}}    
\def\pasa{\refa@jnl{PASA}}               
\def\pasp{\refa@jnl{PASP}}               
\def\pasj{\refa@jnl{PASJ}}               
\def\rmxaa{\refa@jnl{Rev. Mexicana Astron. Astrofis.}}%
\def\qjras{\refa@jnl{QJRAS}}             
\def\skytel{\refa@jnl{S\&T}}             
\def\solphys{\refa@jnl{Sol.~Phys.}}      
\def\sovast{\refa@jnl{Soviet~Ast.}}      
\def\ssr{\refa@jnl{Space~Sci.~Rev.}}     
\def\zap{\refa@jnl{ZAp}}                 
\def\nat{\refa@jnl{Nature}}              
\def\iaucirc{\refa@jnl{IAU~Circ.}}       
\def\aplett{\refa@jnl{Astrophys.~Lett.}} 
\def\apspr{\refa@jnl{Astrophys.~Space~Phys.~Res.}}
\def\bain{\refa@jnl{Bull.~Astron.~Inst.~Netherlands}} 
\def\fcp{\refa@jnl{Fund.~Cosmic~Phys.}}  
\def\gca{\refa@jnl{Geochim.~Cosmochim.~Acta}}   
\def\grl{\refa@jnl{Geophys.~Res.~Lett.}} 
\def\jcp{\refa@jnl{J.~Chem.~Phys.}}      
\def\jgr{\refa@jnl{J.~Geophys.~Res.}}    
\def\jqsrt{\refa@jnl{J.~Quant.~Spec.~Radiat.~Transf.}}
\def\memsai{\refa@jnl{Mem.~Soc.~Astron.~Italiana}}
\def\nphysa{\refa@jnl{Nucl.~Phys.~A}}   
\def\physrep{\refa@jnl{Phys.~Rep.}}   
\def\physscr{\refa@jnl{Phys.~Scr}}   
\def\planss{\refa@jnl{Planet.~Space~Sci.}}   
\def\procspie{\refa@jnl{Proc.~SPIE}}   

\let\astap=\aap
\let\apjlett=\apjl
\let\apjsupp=\apjs
\let\applopt=\ao

\title{MHD Models of Gamma-ray Emission in WR~11}
\author[aff1]{K.~Reitberger\corref{cor1}}

\author[aff1]{R.~Kissmann}
\author[aff2]{A.~Reimer}
\author[aff1]{O.~Reimer}

\affil[aff1]{Institut f\"ur Astro- und Teilchenphysik, Leopold-Franzens-Universit\"at Innsbruck, A-6020 Innsbruck, Austria}
\affil[aff2]{Institut f\"ur Theoretische Physik, Leopold-Franzens-Universit\"at Innsbruck, A-6020 Innsbruck, Austria}

\corresp[cor1]{Corresponding author: klaus.reitberger@uibk.ac.at}

\maketitle

\begin{abstract}
Recent reports claiming tentative association of the massive star binary system $\gamma^2$~Velorum (WR 11) with a high-energy $\gamma$-ray source observed by Fermi-LAT contrast the so-far exclusive role of $\eta$~Carinae as the hitherto only detected $\gamma$-ray emitter in the source class of particle-accelerating colliding-wind binary systems. We aim to shed light on this claim of association by providing dedicated model predictions for the nonthermal photon emission spectrum of WR~11. \\
We use three-dimensional magneto-hydrodynamic modeling to trace the structure and conditions of the wind-collision region of WR~11 throughout its 78.5 day orbit, including the important effect of radiative braking in the stellar winds. A transport equation is then solved in the wind-collision region to determine the population of relativistic electrons and protons which are subsequently used to compute nonthermal photon emission components.\\
We find that -- if WR~11 be indeed confirmed as the responsible object for the observed $\gamma$-ray emission -- its radiation will unavoidably be of hadronic origin owing to the strong radiation fields in the binary system which inhibit the acceleration of electrons to energies sufficiently high for observable inverse Compton radiation. Different conditions in wind-collision region near the apastron and periastron configuration lead to significant variability on orbital time scales. The bulk of the hadronic $\gamma$-ray emission originates at a $\sim$400 R$_\odot$ wide region at the apex.

\end{abstract}

\section{INTRODUCTION}
The surprising discrepancy between the prediction of high-energy $\gamma$-ray emission from colliding wind binaries \citep{Benaglia2003,Reimer2006,Pittard2006} and their ongoing non-detection \citep{Werner2013} is still unresolved. The hitherto only exception to the rule, the bright $\gamma$-ray source $\eta$ Carinae \citep{Tavani2009,Reitberger2015}, renders the lack of emission from similar sources even more interesting.

The recently reported analysis of 7 promising colliding-wind binary systems with 7 years of data from the \textit{Fermi}-LAT by \citet{Pshirkov2016} even adds to the problem by further lowering the upper flux limit of WR~140 -- once predicted to be a promising $\gamma$-source -- to less than 1.1$\times$10$^{-9}$ ph cm$^{-2}$s$^{-1}$ in the 0.1$-$100 GeV energy range. The respective value for $\eta$ Carinae is 184$\pm$30 $\times$10$^{-9}$ ph cm$^{-2}$s$^{-1}$ \citep{3FGL}. Taking into account that the two systems have a similar distance from Earth, similar eccentricities and stellar separations, and even that the total kinetic wind energy available for particle acceleration and ultimately for $\gamma$-ray emission in WR~140 is a factor of 2 larger than for $\eta$~Carinae \citep{Becker2013} - how can this difference in flux of more than two orders of magnitude be explained? The answer might be connected with the many peculiarities of the $\eta$ Carinae system. It is singled out amongst other colliding-wind binary systems by the nature of its primary star, presumably a Luminous Blue Variable with a stellar wind that is ~1.5 orders of magnitude denser then the primary wind of e.g. WR~140.

\citet{Pshirkov2016} now reports the detection of WR~11, a short-period WC8 + O7.5 binary system very unlike the systems discussed above \citep[see e.g.,][]{Becker2013}. However, its closeness to Earth with a distance of merely 340 pc \citep{North2007} and the knowledge we have of its stellar and stellar wind parameters make WR~11 a prime target for numerical modeling. Insights gained by reproducing the observed data with our models may be used to further constrain model parameters needed for the modeling of more complex systems. The ability to successfully model the observed $\gamma$-ray emission will support the claim of association.

By performing three-dimensional magneto-hydrodynamic (MHD) simulation of WR~11, WR~140, and $\eta$ Carinae, we strive to obtain a better understanding of the physics of colliding-wind binary systems, ultimately seeking an explanation for the apparent lack of $\gamma$-ray emission in some of these systems. In this work we present our model results for WR~11.

\section{CHALLENGES OF MODELLING COLLIDING-WIND BINARIES}

\subsection{Model overview}

Following a procedure similar to the one detailed in \citet{Reitberger2014} and \citet{Kissmann2016}, we use the \textsc{Cronos}-code \citep{Kleimann2009} to perform a threedimensional MHD simulation of the colliding-wind binary system. The stellar winds are implemented using a modified Castor-Abbott-Klein (CAK) approximation \citep[based on][]{Pauldrach1986}. In additional to the eight MHD variables (being density $\rho$, velocity $\vec{v}$, temperature $T$, and magnetic field strength $\vec{B}$), we include 200 additional scalar fields representing electrons and protons at 100 energy bins in the range of 1 MeV to 10 TeV. The transport equation is solved in runtime along with the MHD equations and includes the relevant energy loss and gain mechanisms for the particles, as well as energy-dependent diffusion. Simulations also consider the orbital motion of the stars.

This method provides spectra of high-energy electrons and protons for every region of the simulated volume. Following the procedure detailed in \citet{Reitberger2014b} this distribution of particles is then used to compute leptonic and hadronic components of high-energy $\gamma$-ray emission depending on the particles' interaction with the surrounding radiation fields, magnetic fields, and wind plasma density. 

More details on the current mode of operation and the performance of our code -- as it is used to model WR~11 and other colliding-wind binary systems -- will be given in a subsequent publication on model results for the WR~11 system. Here, we restrict ourselves to list a number of major challenges involved in the simulation.

\subsection{Radiative braking}
Radiative braking occurs when the stellar wind in a binary system is efficiently slowed down by the radiative influence of the second star before it reaches the collision zone. There is evidence that this effect is important in WR~11 \citep{Henley2005}, as well as during the periastron passage of $\eta$~Carinae \citep[e.g.,][]{Parkin2009}. The proper implementation of radiative braking is non-trivial as the computation of the force caused by one star's radiation field on the second star's wind is ambiguously treated in literature. There are the two possibilities of either using star-specific CAK-parameters (weak coupling) or wind-specific CAK-parameters (strong coupling). Numerical simulations of CWB-systems have hitherto predominantly used the former method \citep[e.g.,][]{Pittard2009, Parkin2011, Madura2013, Reitberger2014}, whereas theoretical papers on radiative inhibition and braking have generally applied the latter method \citep[e.g.,][]{Stevens1994, Owocki1995, Gayley1997}. We find that in the case of WR~11 sufficient agreement with observations from X-ray spectroscopy can only be achieved by using strong coupling. Therefore we consistently use this method.

\subsection{The magnetic field} 
Care must be taken in the implementation of the magnetic field. The effect that a strong dipole field has on the wind plasma of a single has star has been studied by \citet{ud-Doula2002} in great detail. Its impact on the shape of and conditions at the wind collision region in a binary system were recently investigated by \citet{Kissmann2016}. It is shown that the wind-collision region can be significantly distorted by a strong magnetic field (e.g., B$_\mathrm{surface}$=100~G). Wind velocities towards the poles of the magnetic field are greatly increased. In order not to encounter the aforementioned distortions, we choose a relatively low surface magnetic field of 10~G. Although such a field does not have any significant influence on the wind structure, we find that it does leave a significant imprint on the electron distribution. Where the poles of the magnetic dipole are close to the wind-collision region, synchrotron losses are high and electrons reach lower energies than along the magnetic equator where the field is weak. 

\subsection{Energy-dependent diffusion} 
Whereas maximum electron energies are determined by synchrotron and inverse Compton losses, which usually dominate the acceleration at higher energies, the maximum energies of the protons are determined by diffusion. An approximation often used is the Bohm cutoff, which simply imposes a maximum energy at which the protons leave the shock fronts before being further accelerated. Such an approximation is no longer necessary if an energy-dependent diffusion coefficient is used \citep{Kirk1998}. In our model, we apply a diffusion coefficient of the form $D(E)=D_0E^\delta$. According to literature, the exponent $\delta$ is generally set to $\delta=0.3$ for a Kolmogorov type spectrum and to $\delta=0.5$ for a Kraichnan type turbulence spectrum \citep{Strong2007}. We try both and compare it to the data. $D_0$ remains a free-parameter.

\subsection{Orbital motion}
Although WR~11 has an eccentricity of 0.3 which is comparably low to the value of $\sim$0.9 for $\eta$ Carinae and WR~140, the conditions at periastron and apastron are far from similar. Whereas the flow downstream of the shock is fairly laminar around apastron, turbulence emerges as the system closes in on its periastron passage. It has also been studied to what degree the distortion of the wind-collision region caused by the orbital motion influences the particle distribution and $\gamma$-ray emission. We find no significant difference in the final $\gamma$-ray emission of a simulation that fully considers orbital motion compared to a simulation showing the steady state at ap- or periastron. 

\section{RESULTS OF MODELLING WR~11}
\subsection{Shape of wind collision region in agreement with observations}
In their analysis of high-resolution X-ray spectroscopy of Chandra data, \citet{Henley2005} find evidence for a large shock-cone opening angle. They remark that simulations hitherto failed to reproduce such a feature which might be due to significant radiative braking. In our simulation, including the effects of radiative braking with strong coupling of the CAK parameters, we can indeed reproduce this large-shock cone opening angle, which is most evident close to the periastron passage. 

\begin{figure}
	\setlength{\unitlength}{0.0013\textwidth}
		\begin{picture}(335,321)(0,0)
		\put(0,0){\includegraphics[height=320\unitlength,trim=0cm 0cm 2.4cm 0 cm, clip=true]{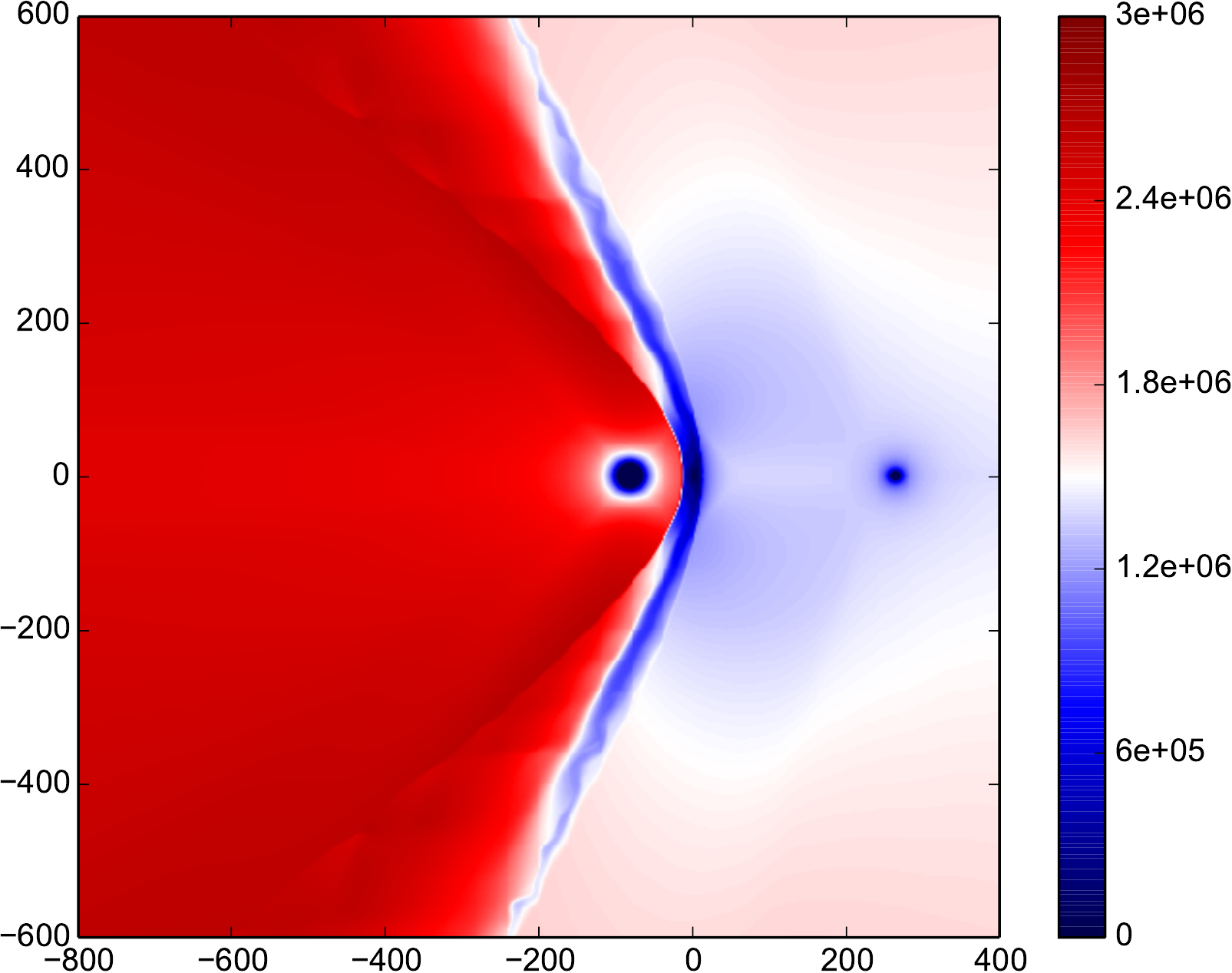}}
		\put(0,150){\rotatebox{90}{\footnotesize $z\;[\mathrm{R}_\odot] $}}
		\put(180,-10){\footnotesize $x\;[\mathrm{R}_\odot] $}
		\put(233,163){\color{yellow}\line(-1,2){80}}
		\put(232.5,163){\color{yellow}\line(-1,2){80}}
		\put(232,163){\color{yellow}\line(-1,2){80}}
		\put(233.5,163){\color{yellow}\line(-1,2){80}}
		\put(233,163){\color{yellow}\line(-1,0){210}}
		\put(233,163.5){\color{yellow}\line(-1,0){210}}
		\put(233,162.5){\color{yellow}\line(-1,0){210}}
		\put(218,166){\color{yellow}$\theta$}
		\end{picture}
		\begin{picture}(335,321)(0,0)
		\put(0,0){\includegraphics[height=320\unitlength,trim=0cm 0cm 2.4cm 0 cm, clip=true]{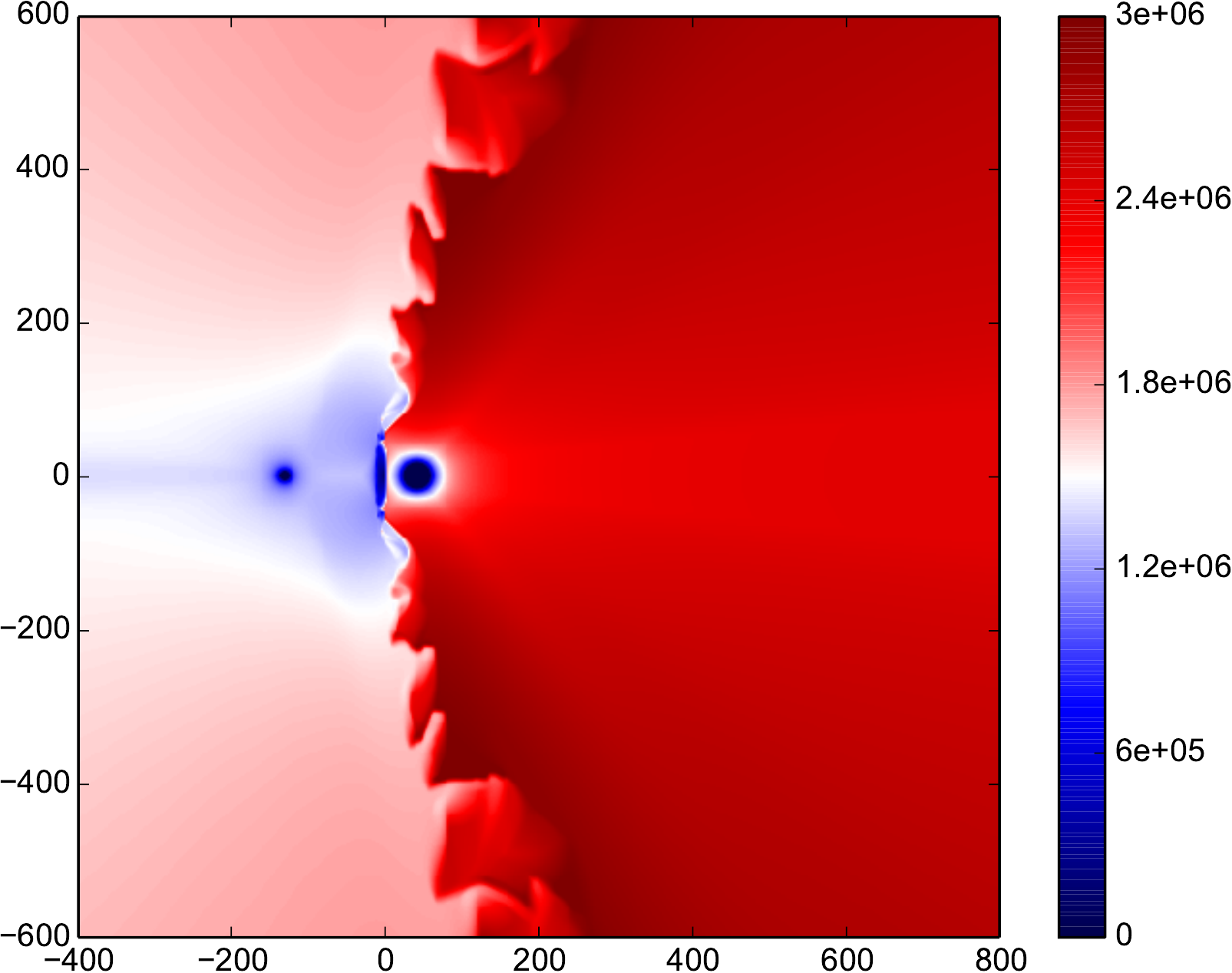}}
		\put(180,-10){\footnotesize $x\;[\mathrm{R}_\odot] $}
		\put(124,163){\color{yellow}\line(1,4){40}}
		\put(124.5,163){\color{yellow}\line(1,4){40}}
		\put(124,163){\color{yellow}\line(1,4){40}}
		\put(124.5,163){\color{yellow}\line(1,4){40}}
		\put(124,163){\color{yellow}\line(1,0){210}}
		\put(124,163.5){\color{yellow}\line(1,0){210}}
		\put(124,162.5){\color{yellow}\line(1,0){210}}
		\put(127,166){\color{yellow}$\theta$}
	\end{picture}
		\begin{picture}(100,321)(0,0)
		\includegraphics[height=320\unitlength,trim=14cm 0cm 0cm 0 cm, clip=true]{pe_vabs_xz.pdf}
					\put(-30,0){$v$ [m s$^{-1}$]}
	\end{picture}\\
	\caption{Absolute velocity of wind plasma for apastron (left) and periastron (right) configuration. The approximate opening angle $\theta$ is indicated by the yellow line.}
	\label{1}
\end{figure}

Figure \ref{1} shows absolute velocity of the wind plasma for the apastron and periastron configuration.The opening angle is also indicated. The strong effect of radiative braking is clearly visible by the blue area upstream of the wind collision where the secondary wind is effectively slowed down by the primary star's radiation. Also, the periastron plot displays the effect of shadowing where the secondary wind is accelerated less in the region where the primary star is eclipsed by the disk of the secondary. As mentioned above, the shock is fairly laminar at apastron and more turbulent at periastron. The shock-cone opening angle is $\sim$65$^\circ$ for apastron and $\sim$76$^\circ$ for periastron.

\subsection{Hadronic dominance in high-energy $\gamma$-ray emission}
\begin{figure}
	\setlength{\unitlength}{0.0008\textwidth}
		\begin{picture}(335,321)(0,0)
		\includegraphics[height=320\unitlength,trim=0cm 0cm 2.2cm 0 cm, clip=true]{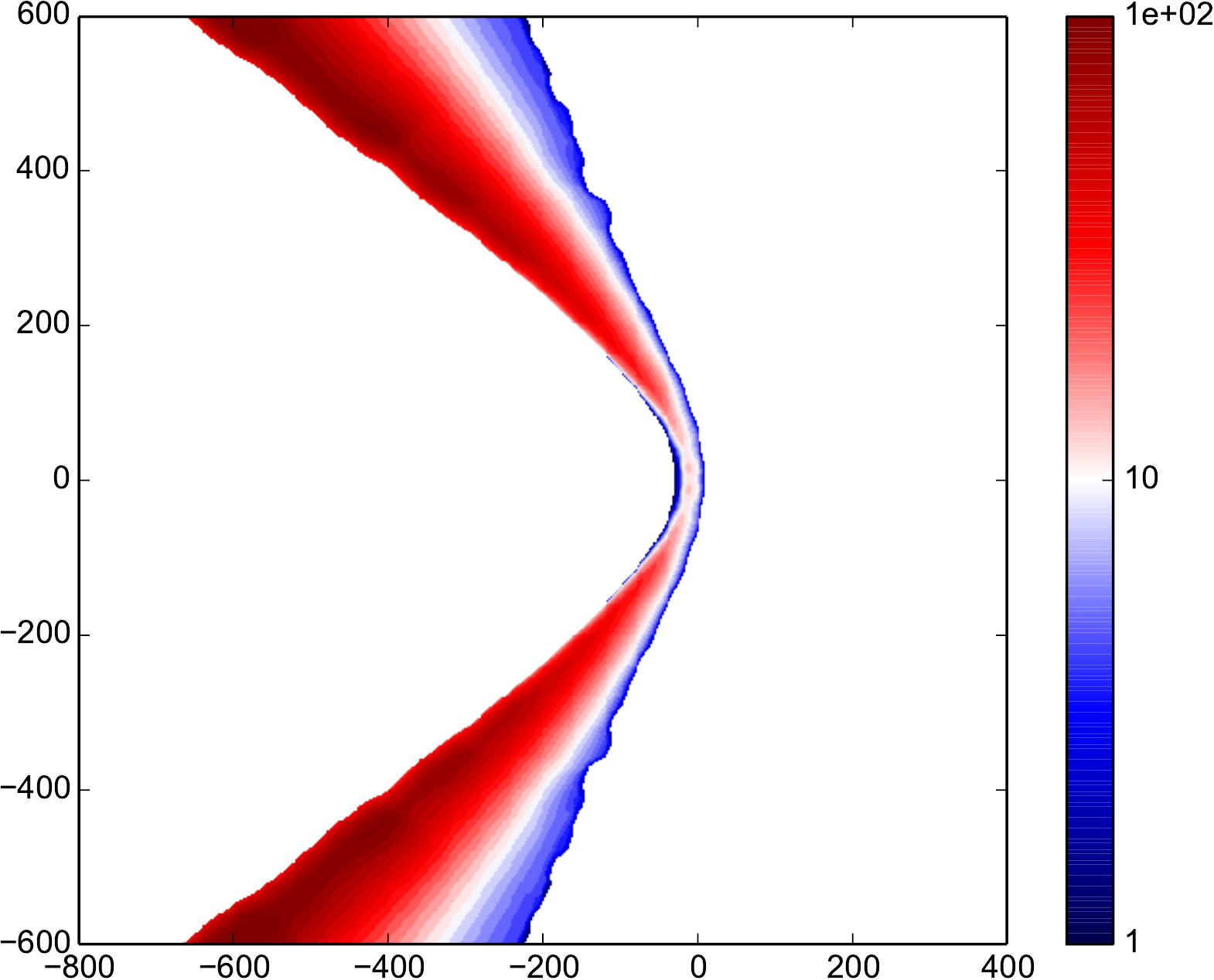}
		\put(-340,150){\rotatebox{90}{\scriptsize $y\;[\mathrm{R}_\odot] $}}
		\put(-190,-20){\scriptsize $x\;[\mathrm{R}_\odot] $}
	\end{picture}	
		\begin{picture}(50,321)(0,0)
		\includegraphics[height=320\unitlength,trim=14cm 0cm 0cm 0 cm, clip=true]{ap_logemax.pdf}
		\put(-60,-15){\scriptsize $E_\mathrm{max}\;[\mathrm{MeV}] $}
	\end{picture}
		\begin{picture}(335,321)(0,0)
		\includegraphics[height=320\unitlength,trim=0cm 0cm 2.2cm 0 cm, clip=true]{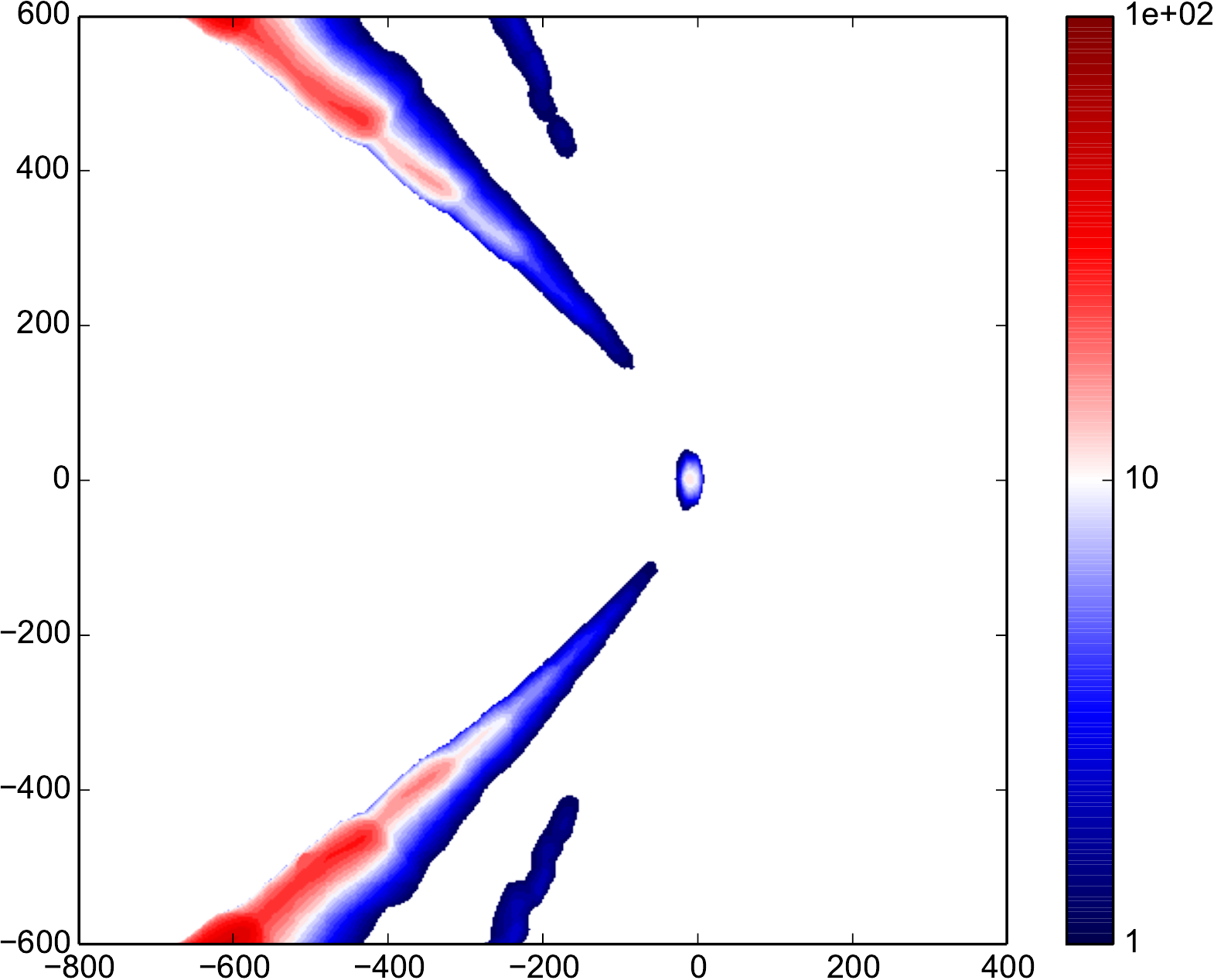}
		\put(-340,150){\rotatebox{90}{\scriptsize $z\;[\mathrm{R}_\odot] $}}
		\put(-190,-20){\scriptsize $x\;[\mathrm{R}_\odot] $}
	\end{picture}		
		\begin{picture}(50,321)(0,0)
		\includegraphics[height=320\unitlength,trim=14cm 0cm 0cm 0 cm, clip=true]{ap_logemax_xz.pdf}
		\put(-60,-15){\scriptsize $E_\mathrm{max}\;[\mathrm{MeV}] $}
	\end{picture}
		\begin{picture}(335,321)(0,0)
		\includegraphics[height=320\unitlength,trim=0cm 0cm 2.2cm 0 cm, clip=true]{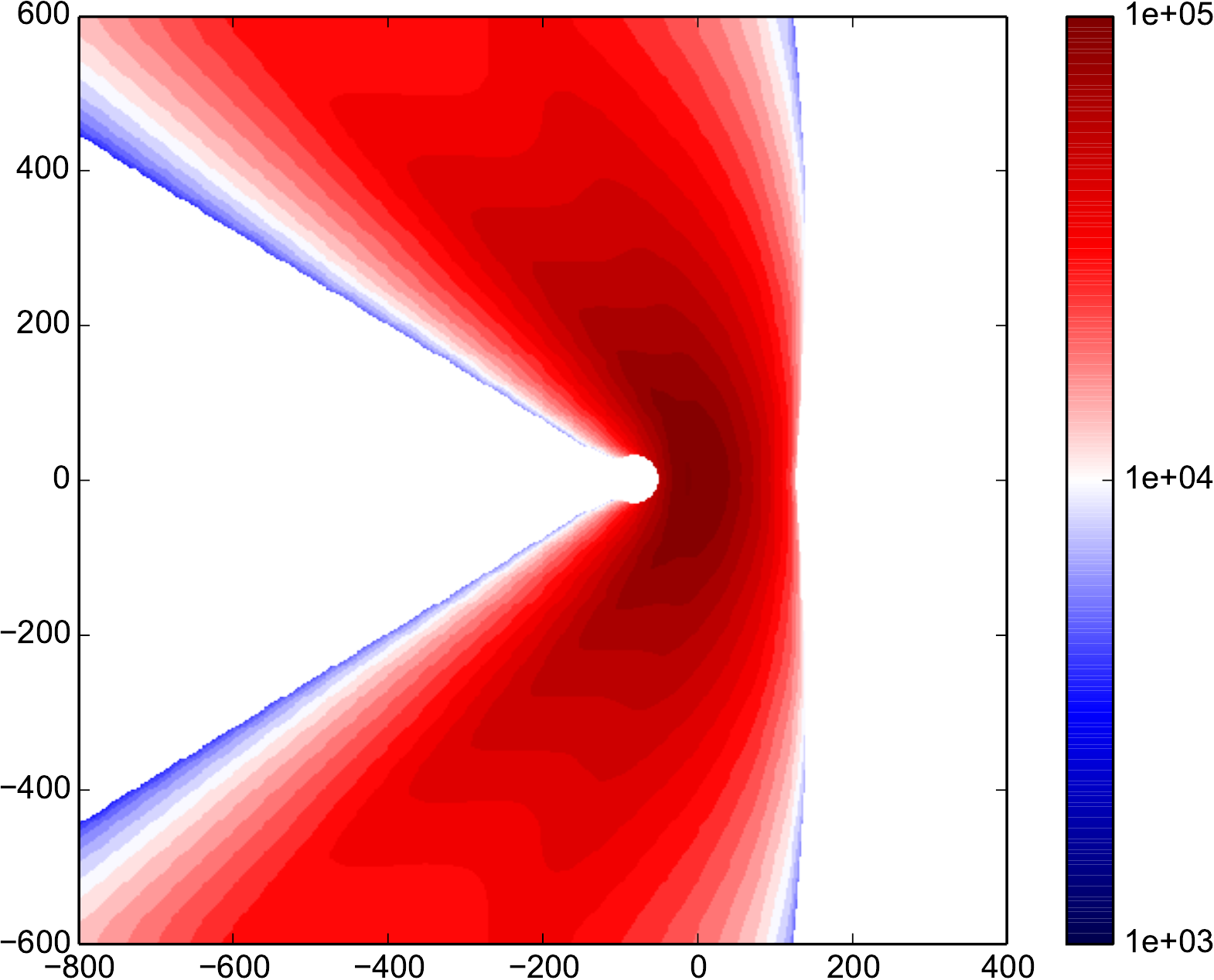}
		\put(-340,150){\rotatebox{90}{\scriptsize $z\;[\mathrm{R}_\odot] $}}
		\put(-190,-20){\scriptsize $x\;[\mathrm{R}_\odot] $}
	\end{picture}	
		
		\begin{picture}(50,321)(0,0)
		\includegraphics[height=320\unitlength,trim=14cm 0cm 0cm 0 cm, clip=true]{ap_logpmax_xz_neu.pdf}
		\put(-66,-15){\scriptsize $E_\mathrm{max}\;[\mathrm{MeV}] $}
	\end{picture}
	\caption{Maximum particle energies for electrons in the $y-z$ plane (left), electrons in the $x-z$ plane (middle) and protons (right).}
	\label{2}
\end{figure}
In a short-period binary system like WR~11, the stellar separation and therefore the distance between the stars and the wind collision region are fairly small. Stellar radiation fields and magnetic fields at the site of the particle acceleration are therefore high compared to long-period binary systems. This leads to severe energy losses by inverse Compton emission and synchrotron emission. We find that the electrons in WR~11 do not reach energies higher than 100 MeV. 

This is shown in the left and center plot of Figure \ref{2} which shows the maximum energies of the electrons in the $x-y$ and $x-z$ plane. The asymmetry is due to the magnetic dipole field which is aligned along the $z$-axis. Electrons reach higher energies along the equator of the dipole where synchrotron losses are low. Highest energies are reached in the outer wings of the the collision region where the distance of the stars increases.

As protons are not affected from similar losses related to the magnetic and radiation fields, they reach energies up to 100~GeV. At these energies the energy-dependent diffusion is strong and allows the accelerated particles to travel upstream outside the collision region. The highest energy protons are found at the apex of the wind collision region. This is shown in the right plot of Figure \ref{2}.

Due to the low maximum energy of electrons, WR~11 is a clear case of hadronic dominance in the high-energy $\gamma$-ray emission.

\subsection{Confinement of emission region to apex of wind-collision}
\begin{figure}
	\setlength{\unitlength}{0.0011\textwidth}
		\begin{picture}(335,321)(50,0)
		\put(0,0){\includegraphics[height=320\unitlength,trim=1.1cm 0.4cm 2.4cm 0 cm, clip=true]{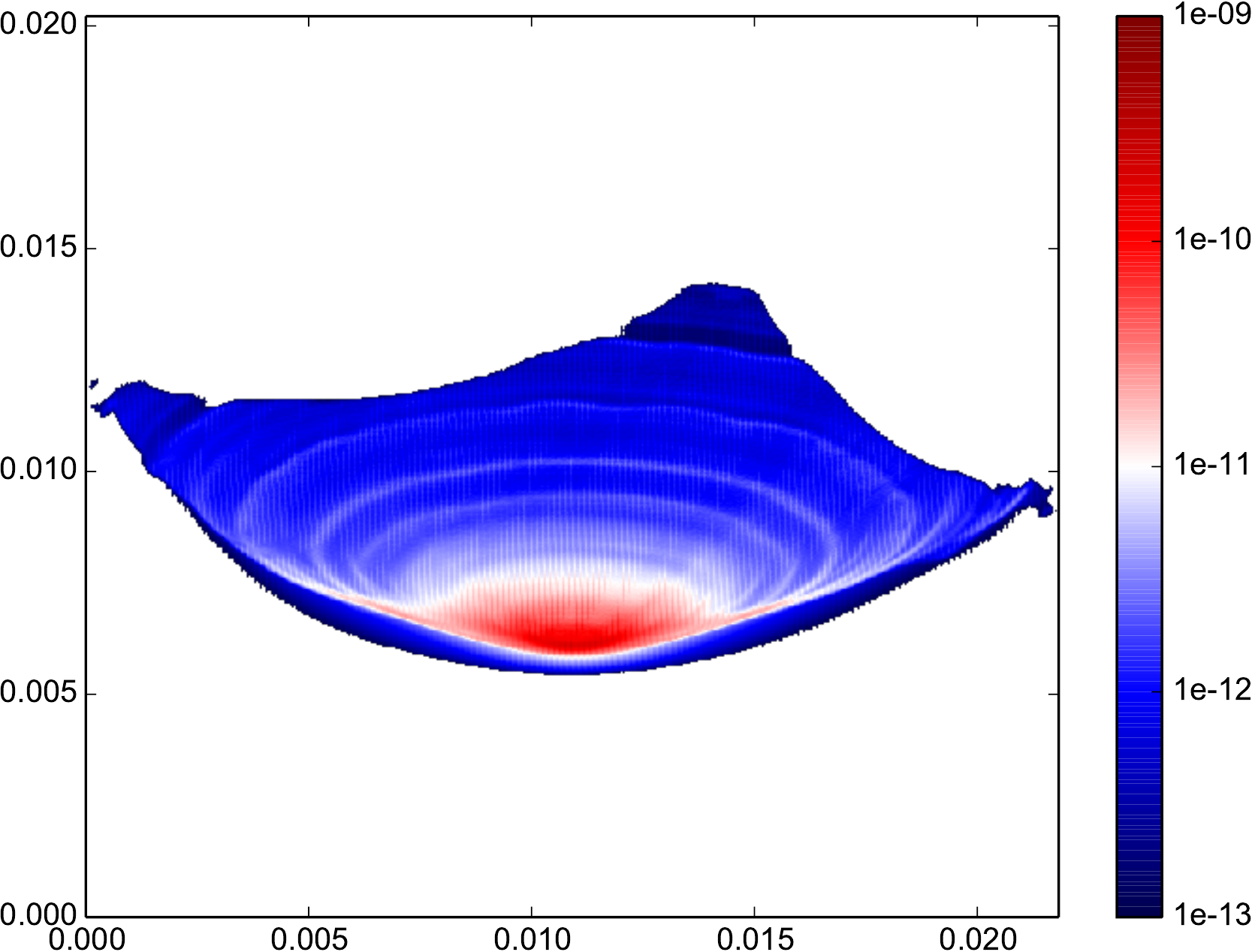}}
		\put(135,20){\line(1,0){79}}
		\put(135,17){\line(0,1){6}}
		\put(214,17){\line(0,1){6}}
		\put(135,25){\scriptsize{0.005 arcsec}}
	\end{picture}	
		\begin{picture}(0,321)(50,0)
		\put(0,0){\includegraphics[height=320\unitlength,trim=14.9cm 0.36cm 0cm 0 cm, clip=true]{p0_2.pdf}}
		\put(-30,-15){\scriptsize Flux [ m$^{-2}$ s$^{-1}$]}
	\end{picture}
	\begin{picture}(335,321)(0,0)
		\put(0,30){\includegraphics[height=280\unitlength,trim=0cm 0cm 0cm 0 cm, clip=true]{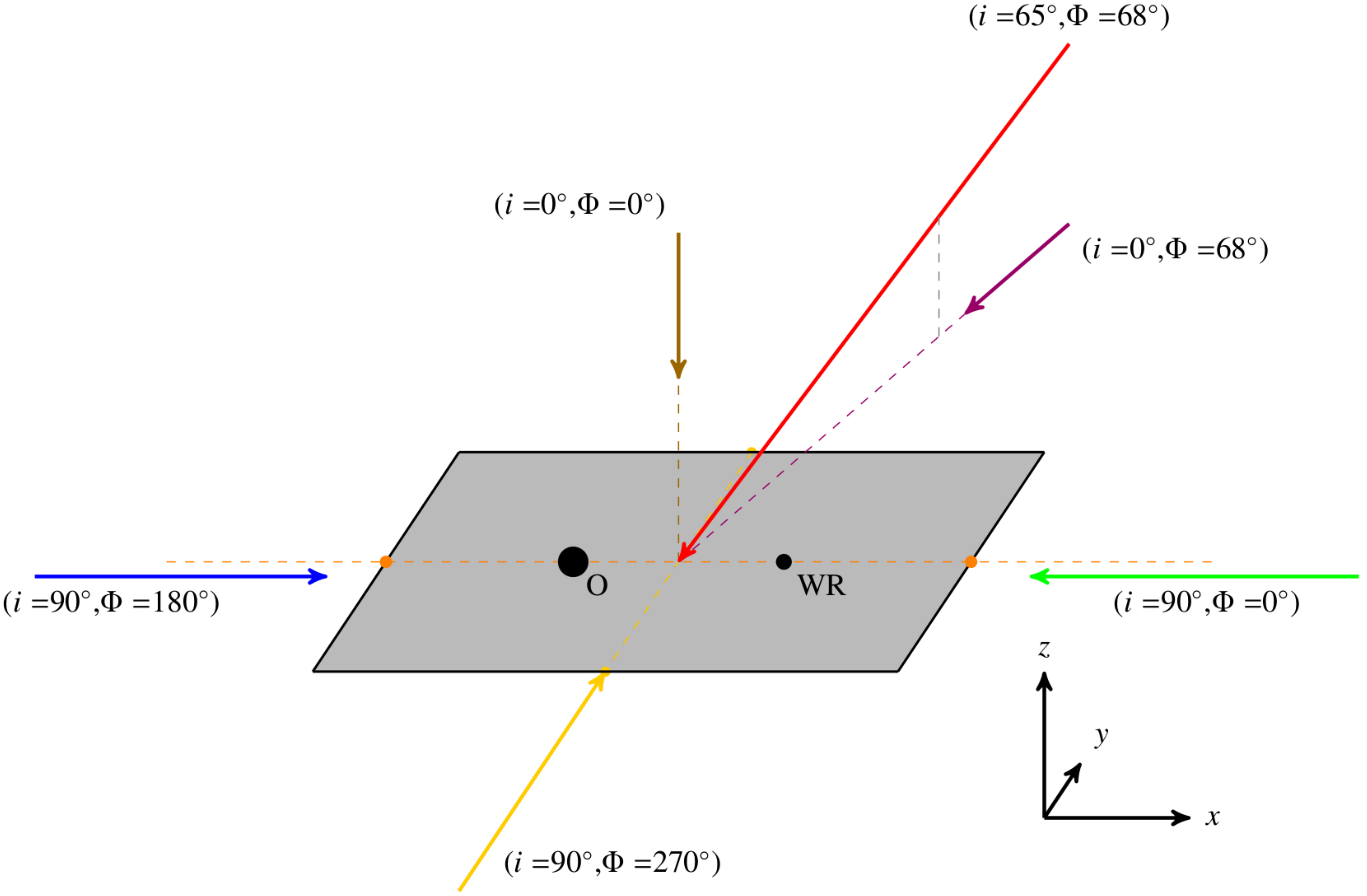}}
	\end{picture}
			\caption{Left: Projected flux above 100 MeV for neutral pion decay. Right: Schematic view of the two stars within the computational domain. The line of centers is represented by the horizontal dashed orange line. Various viewing angles (lines-of-sight) are indicated, including the one that is used in our simulation: $i=65^\circ, \Phi=68^\circ$.}
	\label{3}
\end{figure}
The inclination $i$ of the orbital plan of the $\gamma^2$~Velorum binary system and its argument of apastron $\omega_\mathrm{WR}$ are well constrained. \citet{Schmutz1997} find $i=65^\circ \pm 8^\circ$ and $\omega_\mathrm{WR}=68^\circ \pm 4^\circ$. 
Using this information as well as the distribution of high-energy protons and the wind-plasma density we compute a projected flux map as shown in the left part of Figure \ref{3}. The right part illustrates the orientation of the orbital plane and the stars at apastron configuration with regard to the line of sight. The projected flux map is shown as it would be seen along the red arrow at a distance of $d=$342~$\mathrm{pc}$. It indicates that most of the $\gamma$-ray emission by neutral pion decay has its origin at the apex of the wind-collision in a region roughly 5 milliarcsec or $~400$ $\mathrm{R}_\odot$ wide.

\subsection{Variability on orbital timescales}
\begin{figure}
	\setlength{\unitlength}{0.00065\textwidth}
	\begin{picture}(490,350)(0,0)
		\put(25,0){\includegraphics[height=350\unitlength,trim=0cm 0.0cm 0cm 0 cm, clip=true]{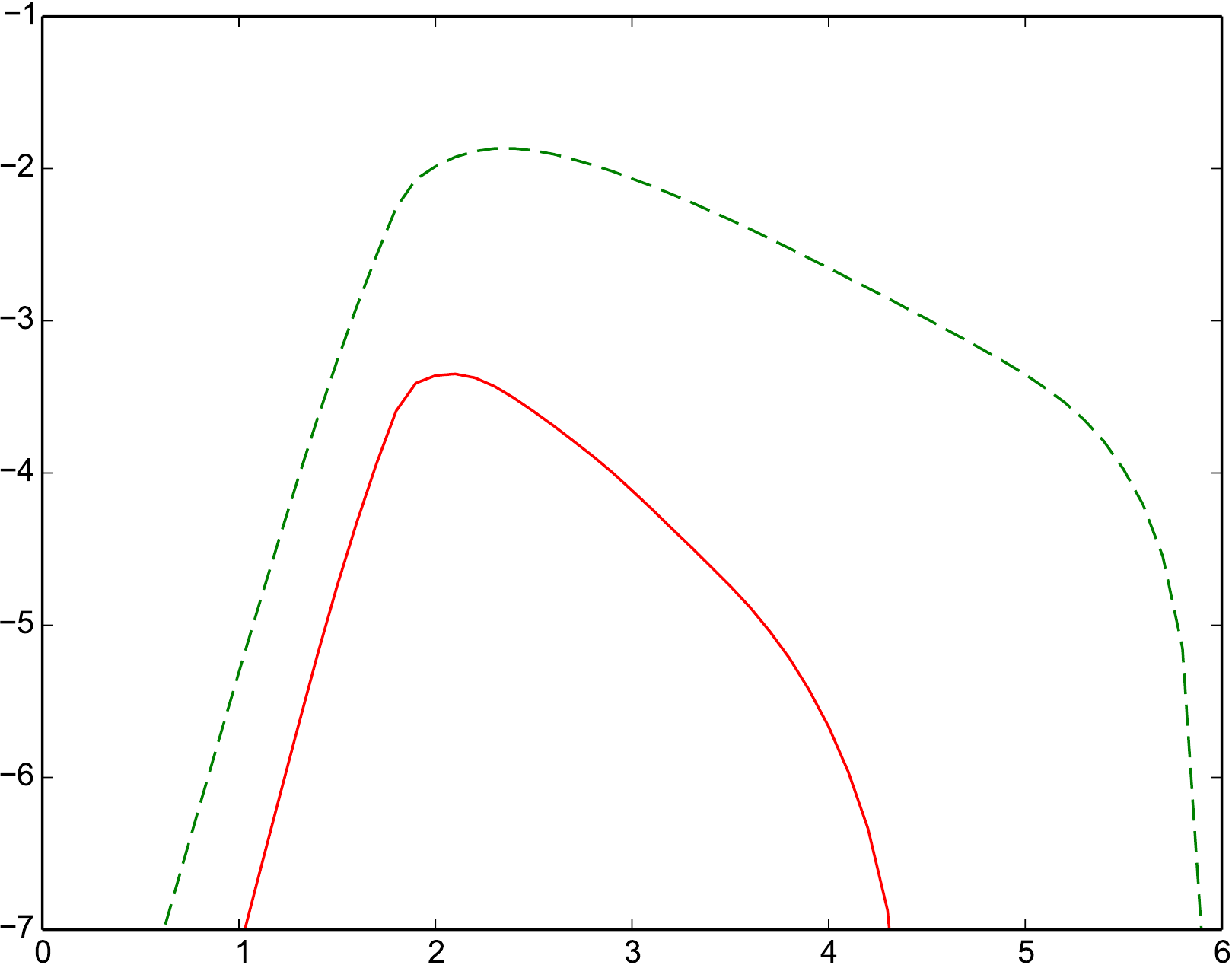}}
		\put(0,80){\rotatebox{90}{\footnotesize $\log$ ( $E^2N$ [MeV m$^{-3}$] ) }}
		\put(150,-20){\footnotesize $\log$ (Energy [MeV])}					
	\end{picture}	
		\begin{picture}(490,350)(0,0)
		\put(20,0){\includegraphics[height=350\unitlength,trim=0cm 0.0cm 0cm 0 cm, clip=true]{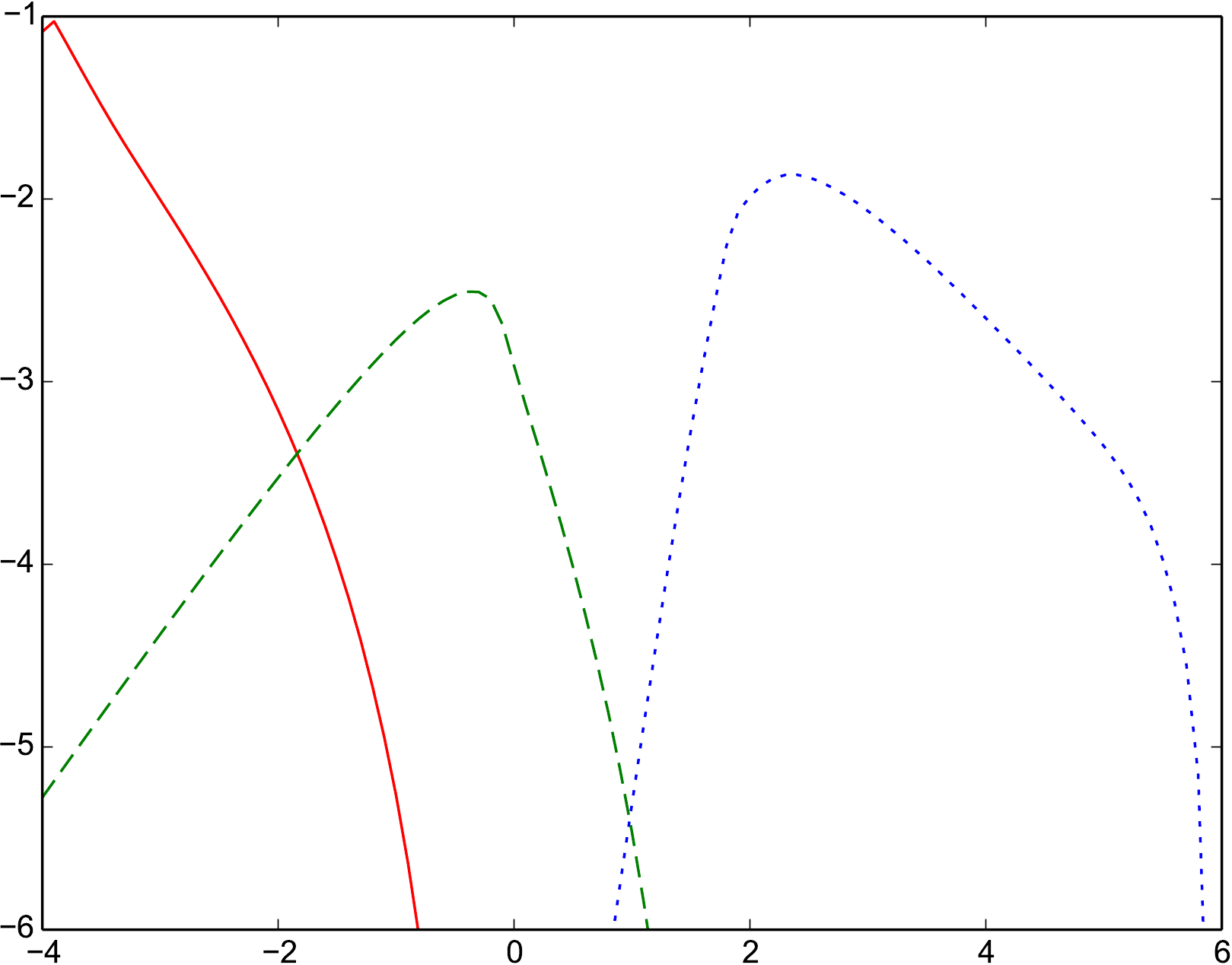}}
		\put(0,80){\rotatebox{90}{\footnotesize $\log$ ( $E^2F$ [MeV m$^{-2}$ s$^{-1}$] ) }}
		\put(150,-20){\footnotesize $\log$ (Energy [MeV])}		
	\end{picture}
	\begin{picture}(490,350)(0,0)
		\put(20,0){\includegraphics[height=350\unitlength,trim=0cm 0.0cm 0cm 0 cm, clip=true]{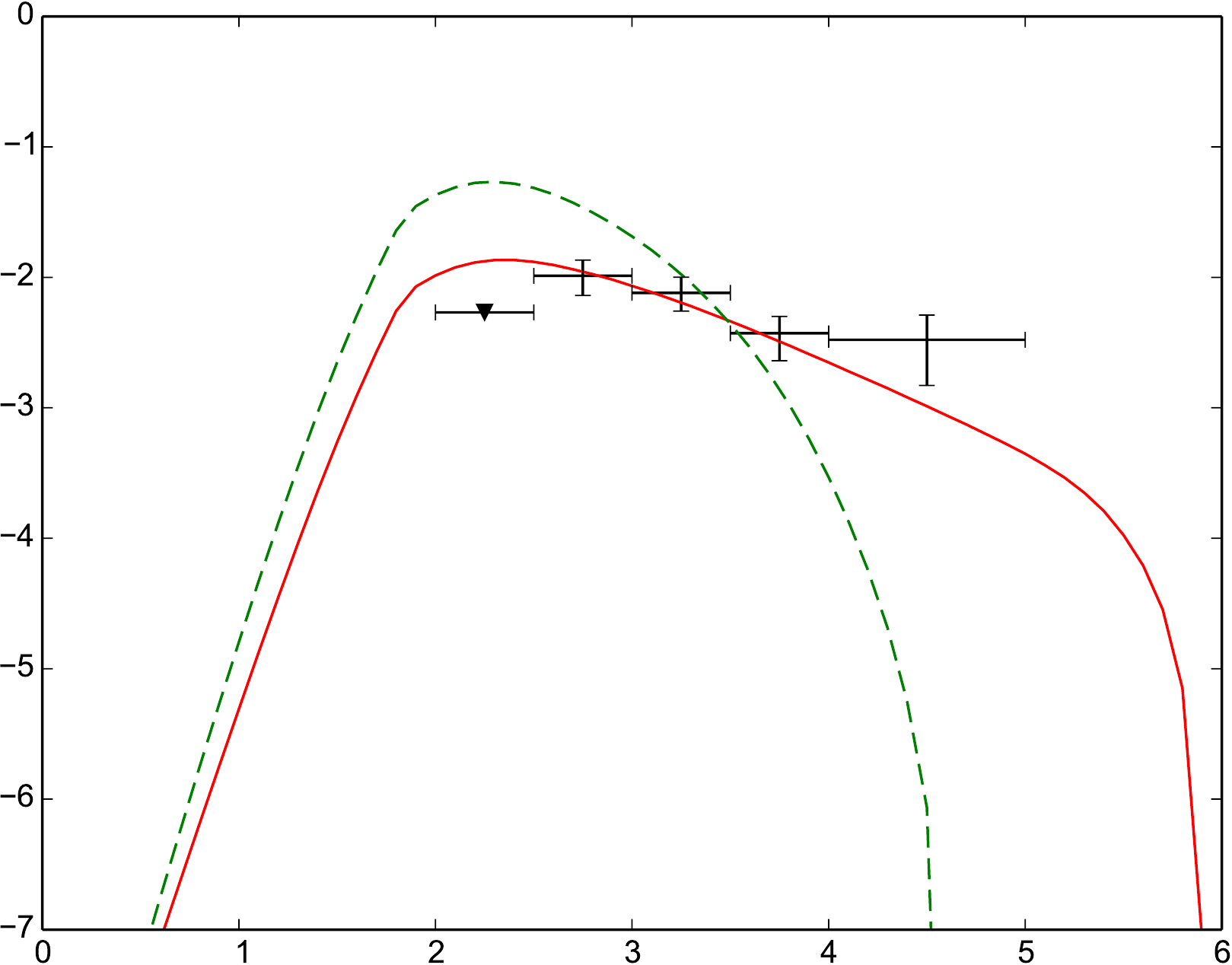}}
		\put(0,80){\rotatebox{90}{\footnotesize $\log$ ( $E^2F$ [MeV m$^{-2}$ s$^{-1}$] ) }}
		\put(150,-20){\footnotesize $\log$ (Energy [MeV])}		
	\end{picture}
		\caption{Left: spectral energy distribution (SED) for neutral pion decay at periastron (red solid) and apastron (green dashed). Middle: SED for inverse Compton emission (red solid), bremsstrahlung (green dashed), and neutral pion decay (blue dotted) at apastron. Right: SED for neutral pion decay with a diffusion exponent $\delta=0.3$ (red solid) and $\delta=0.5$ (green dashed) along with the data points as of \citet{Pshirkov2016}.}
	\label{4}
\end{figure}
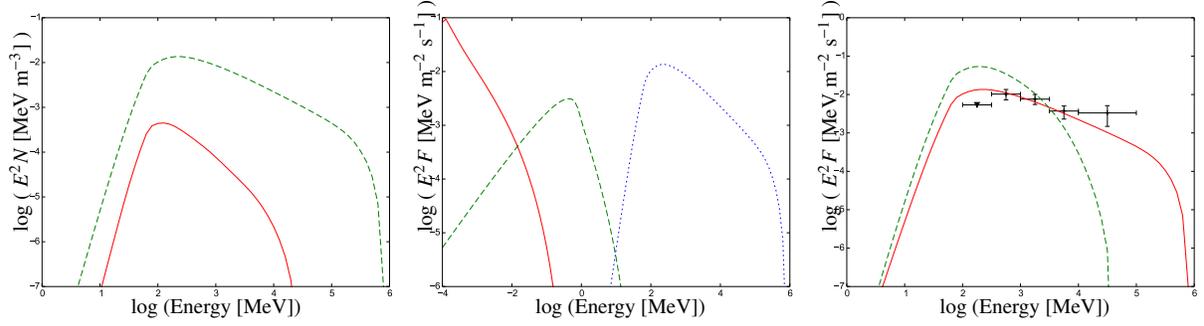

The comparison of the absolute velocities in Figure \ref{1} already suggests that the different conditions at the wind collision region for periastron and apastron configuration will influence particle acceleration and $\gamma$-ray emission. Indeed, we find that the maximum $\gamma$-ray energies as well as the maximum differential flux level differ by almost 2 orders of magnitude. This is shown in the left plot of Figure \ref{4}. Apastron conditions are clearly more favourable for high-energy $\gamma$-ray emission. However, the  very low statistics will make such an expected variability difficult to observe.

\subsection{Broadband spectra}
Although electron energies are far too low to produce high-energy $\gamma$-ray emission we can still use their distribution to compute nonthermal photon emission from inverse Compton scattering and bremsstrahlung emission at lower energies. Thus we compute broadband spectra, as shown in the center plot of Figure \ref{4}, which can also help to constrain important parameters, e.g. limits on the magnetic field strengths.

\subsection{Fitting the diffusion coefficient}
As mentioned in the above discussion of energy-dependent diffusion, the diffusion index $\delta$ can be assumed to have either the value 0.3 (Kolmogorov) or 0.5 (Kraichnan). We apply our model to the data by \citet{Pshirkov2016} and find that the data agrees far better with modelling results for $\delta=0.3$ which is expected for a three dimensional model. Model results for the two values of $\delta$ - all other parameters equal - are shown in the right plot of Figure \ref{4} along with the data.

\section{SUMMARY AND OUTLOOK}
We are using an improved version of a previously presented \citep{Reitberger2014,Reitberger2014b,Kissmann2016} MHD code to simulate the high-energy $\gamma$-ray emission of massive-star binary systems with colliding winds. Three systems, WR~11, $\eta$ Carinae, and WR~140, are of particular interest as the parameters necessary for modelling are sufficiently constrained and observations either yield detection or suprisingly low uper limits. In this work, we present our results for WR~11, showing a) agreement of the wind structure with observations from X-ray spectroscopy, b) hadronic dominance of high-energy $\gamma$-ray emission, c) a well confined emission region close to the apex of the collision region, and d) expected variability in apastron to periastron flux. Our model results support the claim of association of the observed $\gamma$-ray source with the WR~11 system.

In the near future we will present our results on $\eta$ Carinae and WR~140, focusing on finding an explanation for the on-going non-detection of the WR~140 which might be connected to the peculiar role of the primary wind in $\eta$~Carinae and the influence of wind composition on the system's capability for efficient particle acceleration.

\section{ACKNOWLEDGMENTS}
The computational results presented have been achieved (in part) using the HPC infrastructure of the University of Innsbruck. A.R. acknowledge s financial support from the Austrian Science Fund (FWF), project P~24926-N27.

\end{document}